\documentclass[final,3p,times,onecolumn]{elsarticle}
\usepackage[T1]{fontenc}
\usepackage{amsmath,amssymb,amsfonts}
\usepackage{graphicx}
\usepackage{textcomp,nicefrac}
\usepackage{subfigure}
\usepackage{array}
\usepackage{url}
\usepackage[colorlinks=true,linkcolor=blue,citecolor=blue,urlcolor=blue]{hyperref}

\begin{document}

\begin{frontmatter}

\title{Simulation Study of the Design and Spatial Resolution of a Novel AC-coupled Cross Strip LGAD}

\author[label1,label2]{Jingkang Xu}
\author[label2,label3]{Mengzhao Li\corref{cor1}}
\ead{mzli@ihep.ac.cn}
\author[label2,label3]{Zhiliang Hu}
\author[label1]{Tianya Wu\corref{cor2}}
\ead{tianya.wu@ncu.edu.cn}
\author[label2]{Mei Zhao}
\author[label2]{Zhijun Liang}
\author[label2,label3]{Lihua Mo}
\author[label2,label3]{Tianjiao Liang}

\affiliation[label1]{organization={School of Information Engineering, Nanchang University},
            city={Nanchang},
            postcode={330031},
            country={China}}
\affiliation[label2]{organization={Institute of High Energy Physics, Chinese Academy of Sciences},
            city={Beijing},
            postcode={100049},
            country={China}}
\affiliation[label3]{organization={Spallation Neutron Source Science Center},
            city={Dongguan},
            postcode={523803},
            country={China}}

\cortext[cor1]{Corresponding author: Mengzhao Li.}
\cortext[cor2]{Corresponding author: Tianya Wu.}

\begin{abstract}
AC-coupled Low Gain Avalanche Diode (AC-LGAD), evolved from the standard LGAD technology, are silicon detectors characterized by exceptional temporal and spatial resolutions. 
This work proposes a novel sensor named AC-coupled Cross Strip LGAD (CS-LGAD) structure with dual-layered overlapping strip electrodes. It can achieve two-dimensional particle position measurement with low readout density while also inheriting the high temporal resolution of the LGAD.
The sensor structure, doping and current-voltage ($I$-$V$) characteristics of the sensor were investigated through TCAD simulations. Additionally, Monte Carlo simulations were employed to simulate the signal response of CS-LGAD to minimum ionizing particles (MIPs). 
The response of MIPs hitting different positions was simulated, and the reconstruction of particle coordinates (x and y) was achieved based on the sharing of charges on the dual-layer electrodes. The spatial resolution might reach 3.6\% and 5.5\% of the pitch size in the x and y directions, respectively, according to the simulation.
The proposed CS-LGAD is a promising 4D detector with low mass and low readout density.
\end{abstract}

\begin{keyword}
AC-coupled Cross Strip LGAD \sep Silicon Detector \sep Spatial resolution
\end{keyword}

\end{frontmatter}

\renewcommand\floatpagefraction{.75}
\renewcommand\topfraction{.75}
\renewcommand\bottomfraction{.75}
\renewcommand\textfraction{.1}
\setcounter{totalnumber}{50}
\setcounter{topnumber}{50}
\setcounter{bottomnumber}{50}

\section{Introduction}
\label{sec:introduction}
Low Gain Avalanche Diodes (LGADs) are silicon-based sensors characterized by their linear controlled avalanche gain properties. They offer a spatial resolution of 10--50~$\mu$m, depending on pixel size, and a timing resolution of approximately 30~ps, which remains below 50~ps even after a radiation fluence of $2.5 \times 10^{15}$~$n_{eq}/cm^2$~\cite{pellegrini2014technology,sadrozinski20184d,wu2023design}. Based on the traditional PIN structure, LGAD achieves internal gain---typically ranging from 10 to 50---by introducing a highly doped $p^+$ gain layer between the $n$ and $p$ regions~\cite{li2022effects,cartiglia2017beam16ps}. 
Due to the presence of this $p^+$ layer, a high electric field is localized within the gain region under reverse bias. When a charged particle traverses the LGAD sensor, primary electron-hole pairs are generated in the depletion region. As primary electrons drift into the high-field $p^+$ region, they acquire sufficient kinetic energy to impact the silicon lattice, triggering avalanche multiplication~\cite{sze2006physics}.
AC-coupled LGAD (AC-LGAD) represent an innovative sensor architecture derived from the LGAD technology~\cite{tornago2021rsd}. By employing a continuous $n^+$ layer (resistive layer), AC-LGAD effectively overcome the fill-factor limitations inherent in conventional LGADs~\cite{mandurrino2020analysis,giacomini2019fabrication,li2023performance}. As a leading technology for 4D detectors, AC-LGAD are expected to provide precise spatio-temporal measurements for future facilities such as the Circular Electron-Positron Collider (CEPC) and the Electron-Ion Collider (EIC), and are also instrumental in dark matter research~\cite{cepcgroup2018conceptual,abdulkhalek2021science,kramberger2021lgad}.

Currently, single-layer strip and pixel electrodes are the main forms of AC-LGAD. 
Although pixel-type AC-LGAD can provide excellent timing resolution of $\sim$20~ps, and 2D spatial resolution varies from 20--70~$\mu$m depending on the hit position, with a high readout channel density ($N^2$), which leads to issues of high power consumption and high cost. 
The strip-type AC-LGAD has a lower readout density, but it can only provide 1D spatial resolution, achieving a timing resolution of $\sim$37~ps, and a spatial resolution of 8--10~$\mu$m for a 80~$\mu$m pitch size~\cite{dutta2025strip,sun2024strip,bishop2024long}. 
To achieve 2D spatial resolution, two layers of detectors are required.

Therefore, this work proposes a novel LGAD named AC-coupled Cross Strip LGAD (CS-LGAD). 
It features dual-layer strip electrodes that overlap each other, separated by a dielectric layer to enable 2D spatial resolution. 
Compared to pixel-type AC-LGAD, this architecture significantly reduces the readout channel density. This paper details the sensor's design, where the sensor structure, doping profiles, and $I$-$V$ characteristics were characterized using Technology Computer-Aided Design (TCAD) simulations to evaluate electrical performance. 
Furthermore, signal responses to Minimum Ionizing Particles (MIPs) were emulated by Monte Carlo methods. 
Finally, the spatial resolution of CS-LGAD under different noise levels is systematically investigated.

\section{Introduction to the CS-LGAD}
\label{sec:design}
Fig. 1(a) shows a cross-sectional view of a AC-coupled CS-LGAD structure with a unit cell ($2 \times 2$ array), featuring two layers of overlapping strip electrodes.CS-LGAD can be designed with multiple strip-shaped electrodes for engineering applications. In this work, we take the sensor with one unit cell as an example for introduction and research.
The strip electrodes are insulated from each other and directly also from the $n^+$ layer. Fig. 1(b) shows the electrode structure. When MIPs hit the sensor, the signal is read out through the coupling capacitance between the strip electrode and the $n^+$ layer. In the X or Y direction, the closer the strip electrode is to the hit position, the larger the signal, thereby enabling the reconstruction of the particle's two-dimensional coordinates.

\begin{table}[htbp]
\centering
\caption{The structure and parameters of the AC-coupled CS-LGAD sensor.}
\begin{tabular}{| >{\centering\arraybackslash}m{1.8cm} | >{\centering\arraybackslash}m{1.8cm} | >{\centering\arraybackslash}m{2.6cm} |}
        \hline
              Name & Imp/Material & Thickness $[\mu\text{m}]$    \\
              \hline
         Substrate   &  Boron  & 1 \\ 
        \hline
        Epitaxial  & Boron & 50   \\ \hline
        $p^+$ Layer  & Boron & 4  \\ \hline
        $n^+$ Layer & Phosphorus & 1  \\ \hline
        JTE & Phosphorus & 7  \\ \hline
        Dielectric & $\text{Si}_3\text{N}_4$ & 0.15  \\ \hline
        AC strips & Aluminum & 0.15  \\ \hline
        
    \end{tabular}
    \label{tab:exTable1}
\end{table}

\begin{figure}[t]
    \centering
    \subfigure[]{
        \includegraphics[width=0.48\textwidth]{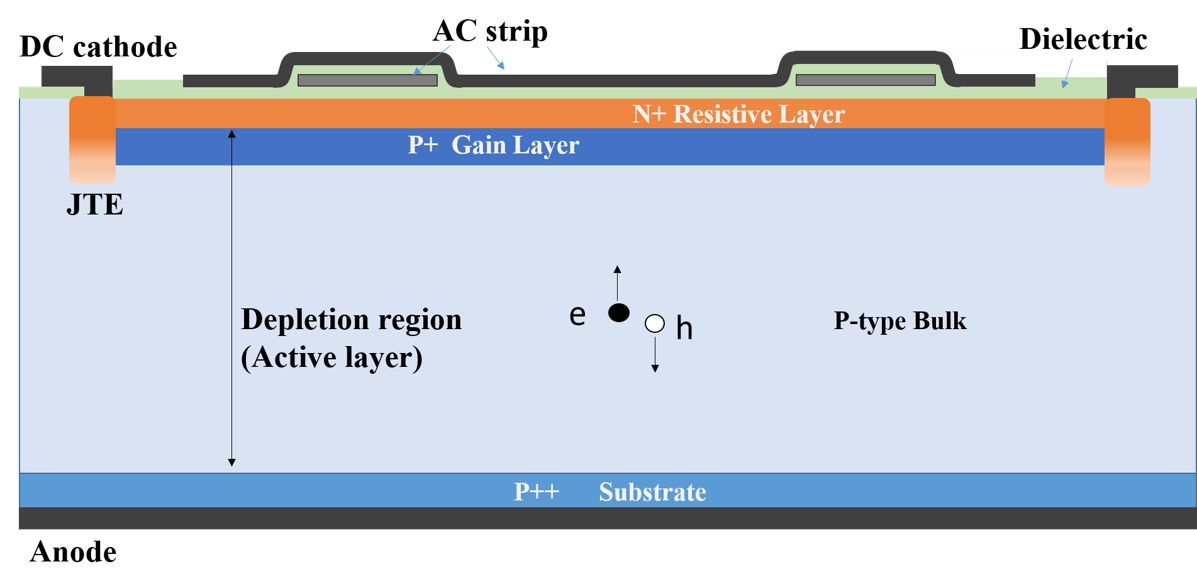}
    } \\
    \subfigure[]{
        \includegraphics[width=0.48\textwidth]{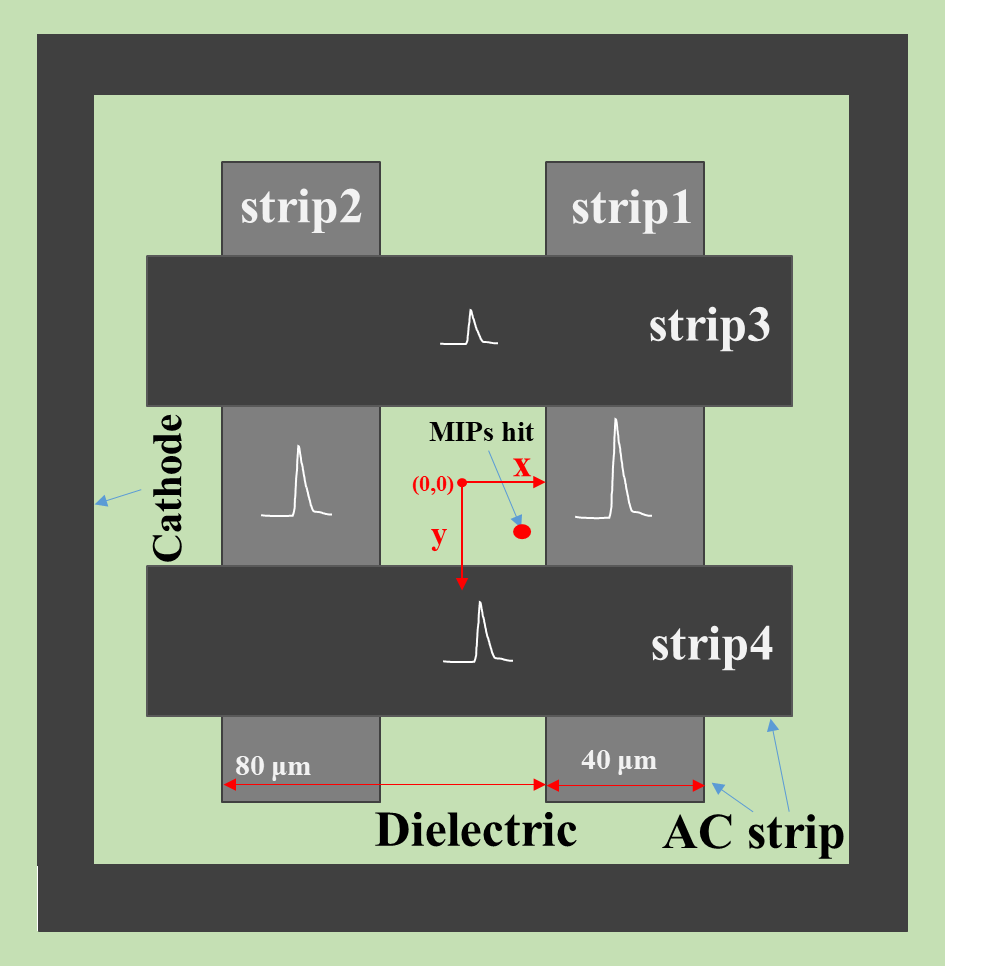}
    }
    \caption{\label{fig:Schematic}(a) Cross-sectional schematic of AC-coupled CS-LGAD sensor with a unit cell.(b) Top view schematic of AC-coupled CS-LGAD sensor with a unit cell.}
\end{figure}

\begin{figure}[t]
    \centering
    \includegraphics[width=0.48\textwidth]{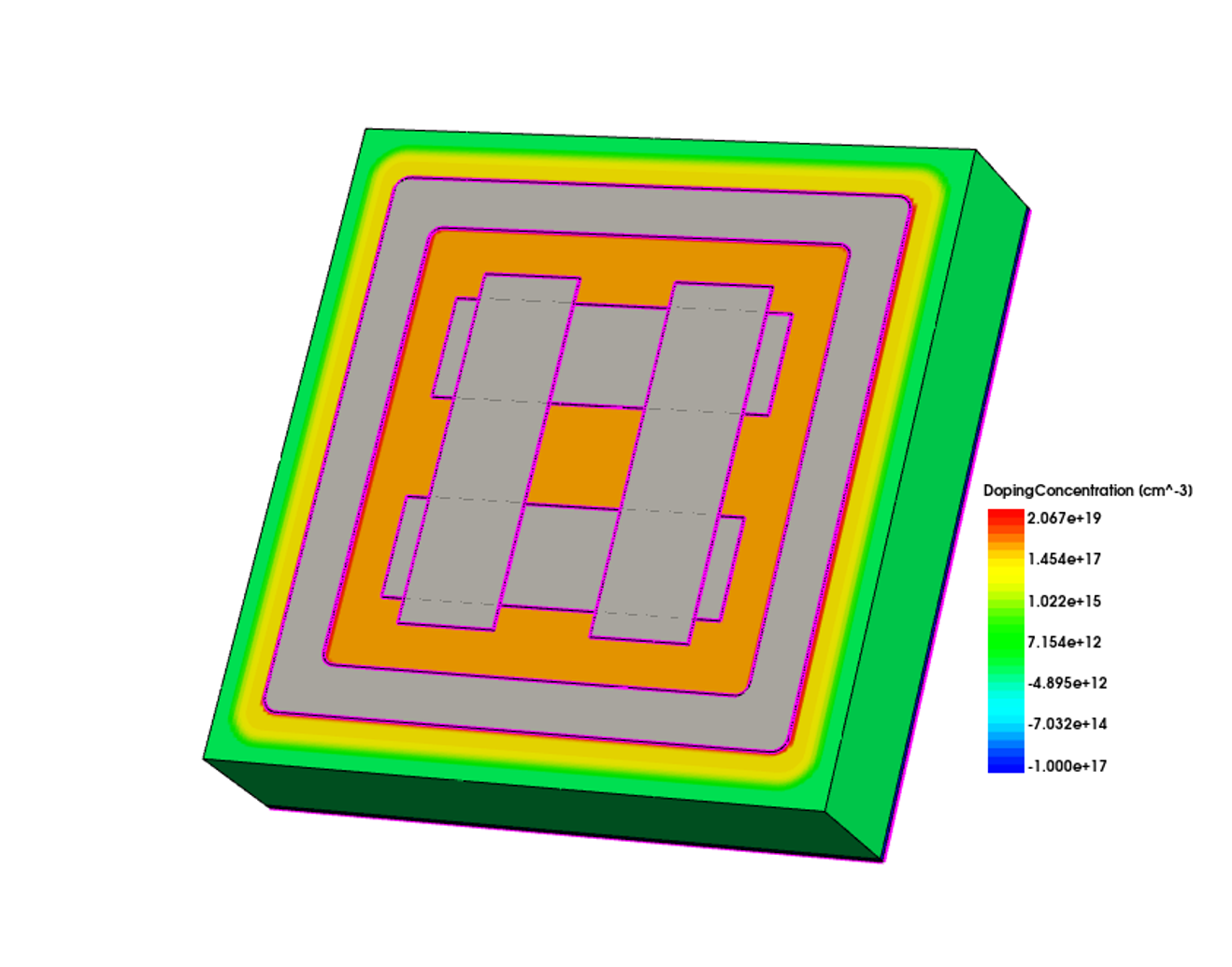}
    \caption{\label{fig:doping}3D structure and doping profile of the AC-coupled CS-LGAD sensor.}
\end{figure}

\begin{figure}[t]
    \centering
    \includegraphics[width=0.48\textwidth]{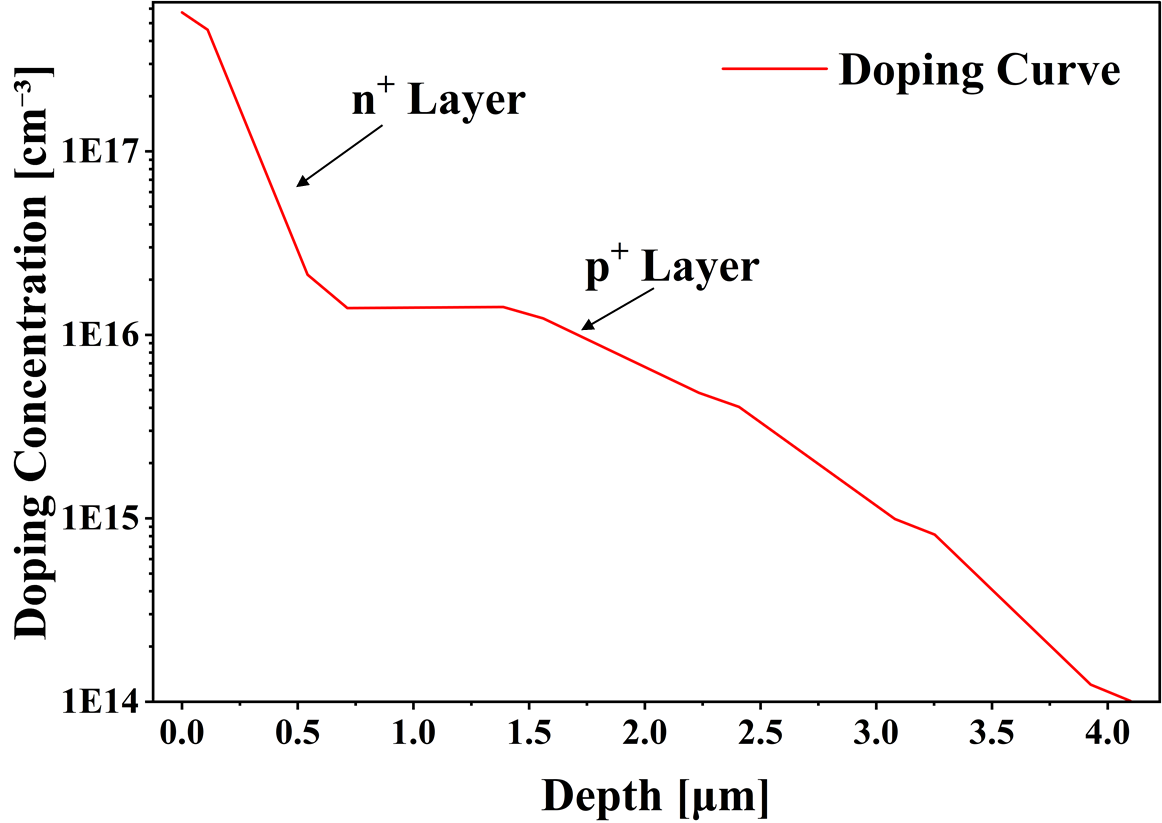}
    \caption{\label{fig:new_figure}The doping concentration curve of the $n^+$ and $p^+$ layer.}
\end{figure}

The 3D architecture and doping profiles of the CS-LGAD, obtained via TCAD simulations, are illustrated in Fig. 2.
The sensor size is $255\,\mu\text{m} \times 255\,\mu\text{m}$ with a $50\,\mu\text{m}$ thick epitaxial layer. The principal structural parameters are summarized in Table 1.
A Junction Termination Extension (JTE) with a phosphorus-doped junction depth of $7\,\mu\text{m}$ surrounds the sensor periphery to alleviate electric field concentration and prevent premature breakdown. 
The size of each AC strip electrode is $40\,\mu\text{m} \times 140\,\mu\text{m}$. 
The sensor incorporates a dual-layer electrode configuration: Strips 1 and 2 comprise the bottom layer, while Strips 3 and 4 form the top layer, separated by a dielectric medium to ensure electrical isolation, with both strip electrodes directly reading signals through their respective strip pads. These  strip electrodes pitch size are $80\,\mu\text{m}$. 
To enhance the coupling capability of the signal, silicon nitride ($\text{Si}_3\text{N}_4$) was selected as the dielectric material instead of silicon dioxide ($\text{SiO}_2$) due to its higher dielectric constant. 
As shown in Fig. 3, the $n^+$ layer exhibits a Gaussian doping profile with a maximum peak concentration of $6 \times 10^{17}\,\text{cm}^{-3}$.
While the total nitride layer thickness is $1\,\mu\text{m}$, it is thinned to $0.15\,\mu\text{m}$ in the induction regions beneath the AC strips to optimize signal coupling. Outside the overlap regions, the dielectric thickness for Strips 3 and 4 remains consistent with that of Strips 1 and 2.

The horizontal Strips 1 and 2 provide $x$-coordinate information, whereas the vertical Strips 3 and 4 facilitate $y$-coordinate readout. Particle incidence occurs within a central $40\,\mu\text{m} \times 40\,\mu\text{m}$ window. Impact events within this region trigger varied amplitude responses across the four AC strips through a charge-sharing mechanism, allowing for precise hit position reconstruction via specialized algorithms.

A metal anode is positioned at the bottom of the sensor to provide the necessary high-voltage bias. Directly above the anode is a $1\,\mu\text{m}$ thick substrate, which is heavily $p^+$-doped with boron to ensure a high-quality Ohmic contact with the aluminum metallization. A $50\,\mu\text{m}$ thick epitaxial (EPI) layer is grown on the substrate, also doped with boron at a high resistivity of $1\,\text{k}\Omega\cdot\text{cm}$ to prevent premature breakdown. This EPI layer serves as the primary drift region for charge carriers generated by particle ionization. Atop the EPI layer, a gain layer is formed using a Gaussian boron doping profile with a diffusion depth of $4\,\mu\text{m}$. This $p$-type gain layer establishes a high electric field region beneath the $n^+$ layer, facilitating the characteristic low-gain properties of the sensor. The $n^+$ layer is phosphorus-doped with a Gaussian profile and a depth of $1\,\mu\text{m}$. This continuous $n$-type resistive layer allows the locally induced signals to undergo lateral diffusion and charge sharing, enabling signal collection by the four overlying AC strip electrodes.

\section{Simulation of I-V and MIPs signals for CS-LGAD}
The current-voltage ($I$-$V$) characteristic is a critical metric for evaluating the static electrical performance of AC-LGAD sensors. It is primarily used to extract the breakdown voltage and evaluate the static electrical behavior of the CS-LGAD sensor.\ At low bias voltages, the leakage current remains relatively stable with negligible variation. As the bias voltage continues to increase, the leakage current rises accordingly. Upon reaching the breakdown threshold, the leakage current surges to the microampere ($\mu$A) level, at which point the sensor can no longer function normally. Fig. 4 shows the simulated $I$-$V$ curve of the CS-LGAD at room temperature (293.15 K), showing a breakdown voltage of approximately -150 V.

\begin{figure}[t]
\centering
\includegraphics[width=0.48\textwidth]{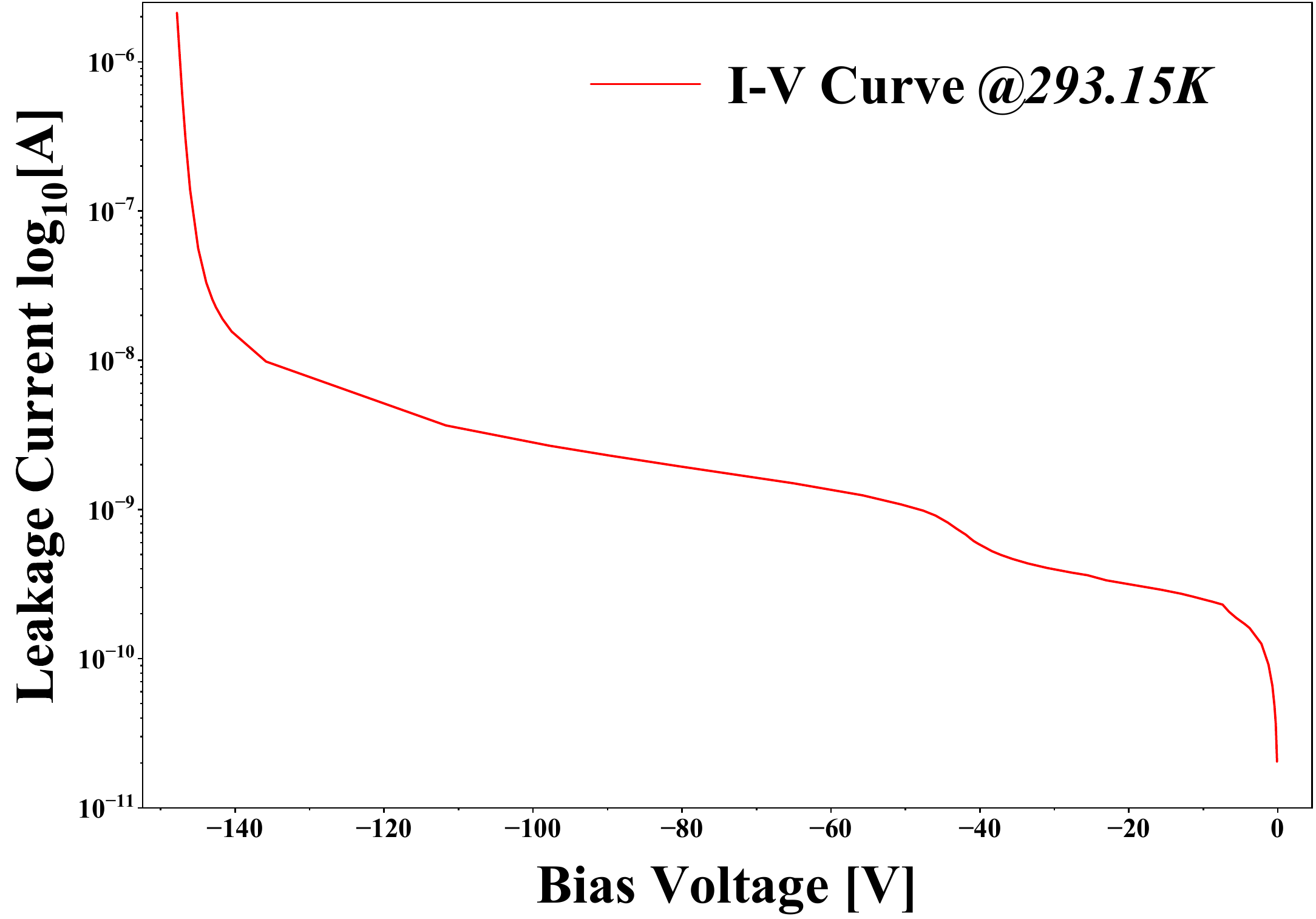}
\caption{\label{fig:iv_curve}$I$-$V$ curve for AC-coupled CS-LGAD.}
\end{figure}

\begin{figure}[t]
\centering
\includegraphics[width=0.48\textwidth]{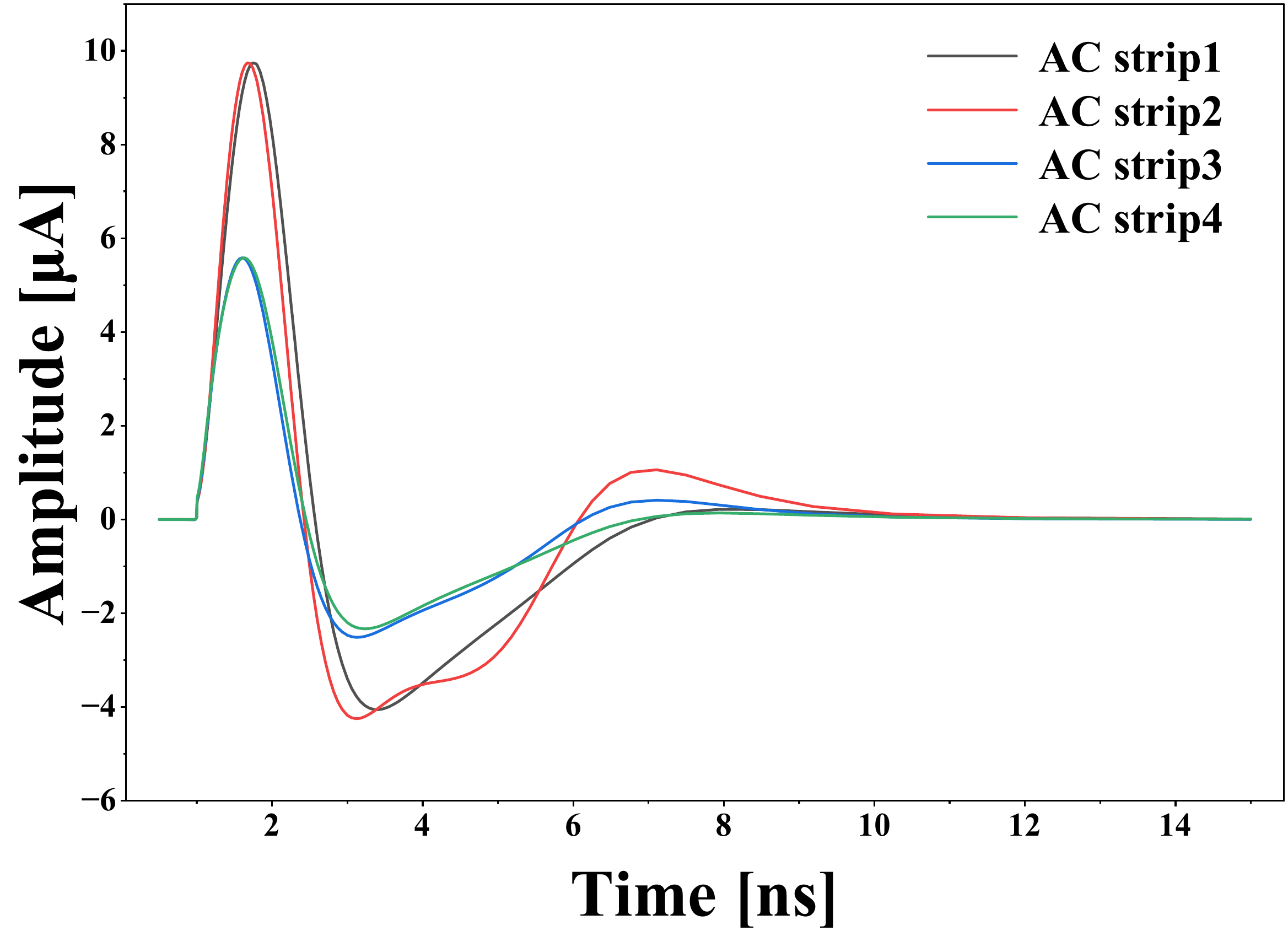}
\caption{\label{fig:signal_response}Signal response to MIPs incidence at position (0,0).}
\end{figure}

\begin{figure}[t]
\centering
\includegraphics[width=0.48\textwidth]{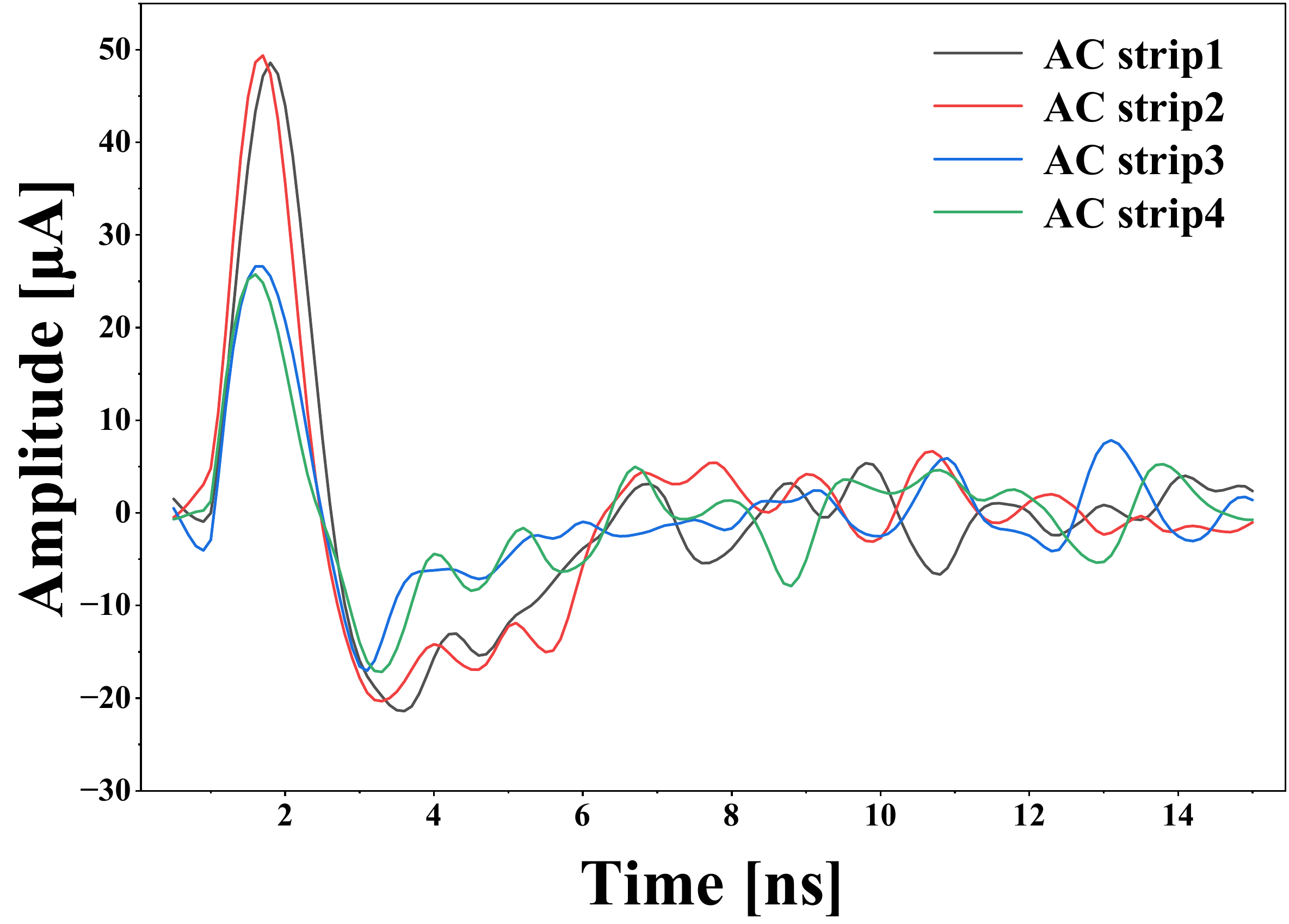}
\caption{\label{fig:noise_signal}Signal waveform after two-stage amplification at position(0,0).}
\end{figure}

\begin{figure}[t]
    \centering
    \includegraphics[width=0.48\textwidth]{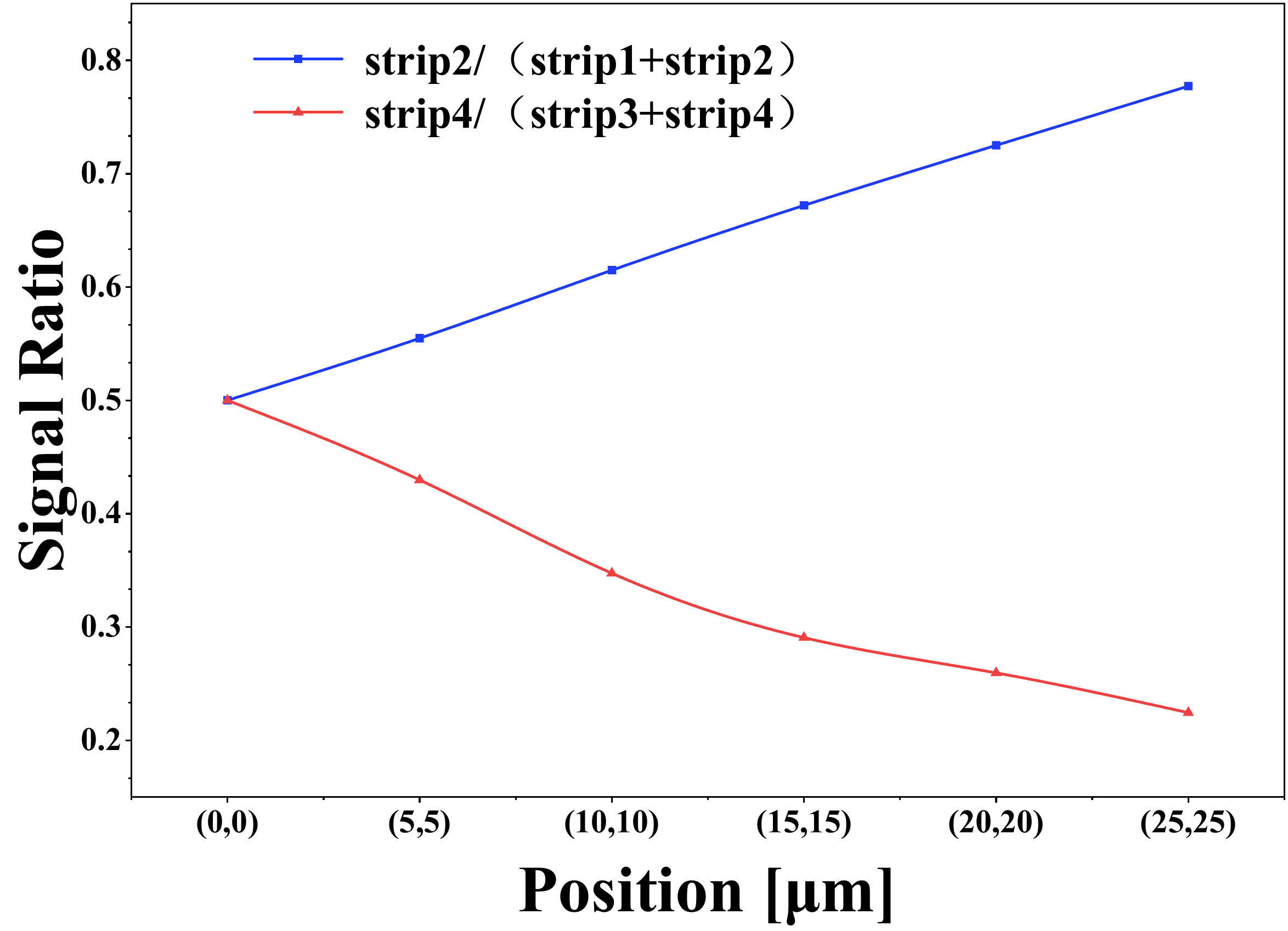}
    \caption{\label{fig:signal_ratio}Signal ratio $R$ as a function of the hit position.}
\end{figure}

Fig. 5 shows the signals of four strip electrodes when a MIP hits the position (0, 0).
When a particle hits the central region between the electrodes, the signal will sharing by the four AC strips in varying proportions. 
Especially, strips 1 and 2 are buried beneath strips 3 and 4; their closer proximity to the $n^+$ layer (which has a thickness of $1\,\mu\text{m}$) results in a higher coupling capacitance. 
Consequently, for a central hit at $(0, 0)$, the majority of the signal charge is induced on the lower strips (1 and 2), leading to higher signal amplitudes compared to those of strips 3 and 4. 
Fig. 5 also shows since the hit occurs at the exact center, the simulated signals for strips 1 and 2 are nearly identical, as are those for strips 3 and 4.
Additionally, a prolonged and complex undershoot is observed in the trailing edge of all signals. This is an inherent feature of AC-LGADs, controlled by the lateral charge diffusion within the resistive $n^+$ layer and the discharge through the distributed RC network.

\section{Position reconstruction methodologies of CS-LGAD}
\label{analysis}
To simulate more realistic experimental conditions, controllable current noise was added to the original current signal and amplified using a two-stage amplifier.
The pre-amplifier is a 470\,$\Omega$ transimpedance amplifier, and the main amplifier is a 20 dB broadband amplifier. Next, the processed signal is used for the reconstruction of the particle hit position.
To ensure the simulation results closely approximate practical experimental conditions, electronic noise was incorporated into each amplification stage and superimposed onto the raw current signal noise. 

Fig. 6 shows the signals of four AC strip electrodes after adding noise and amplification. Based on the noise level of IHEP AC-LGAD Beta source test, a noise Root Mean Square (RMS) value of $3.5\,\text{mV}$ was selected for the simulation to approximate the actual test, with a Signal-to-Noise Ratio (SNR $\approx 14$) for the CS-LGAD sensor.

Spatial resolution is an important parameter of the CS-LGAD, which can be studied by reconstructing the particle position from the signals of the four AC strip electrodes.
Reconstruction methods rely on the variation in signal amplitudes among adjacent AC strip electrodes relative to the particle impact position. When a particle traverses the inter-strip region, the signal amplitudes of strip1 and strip2 (representing the $x$-direction) and strip3 and strip4 (representing the $y$-direction) fluctuate as a function of the hit location. The signal ratio formula for position estimation is defined as follows:

\begin{figure}[t]
\centering
\subfigure[]{
    \includegraphics[width=0.48\textwidth]{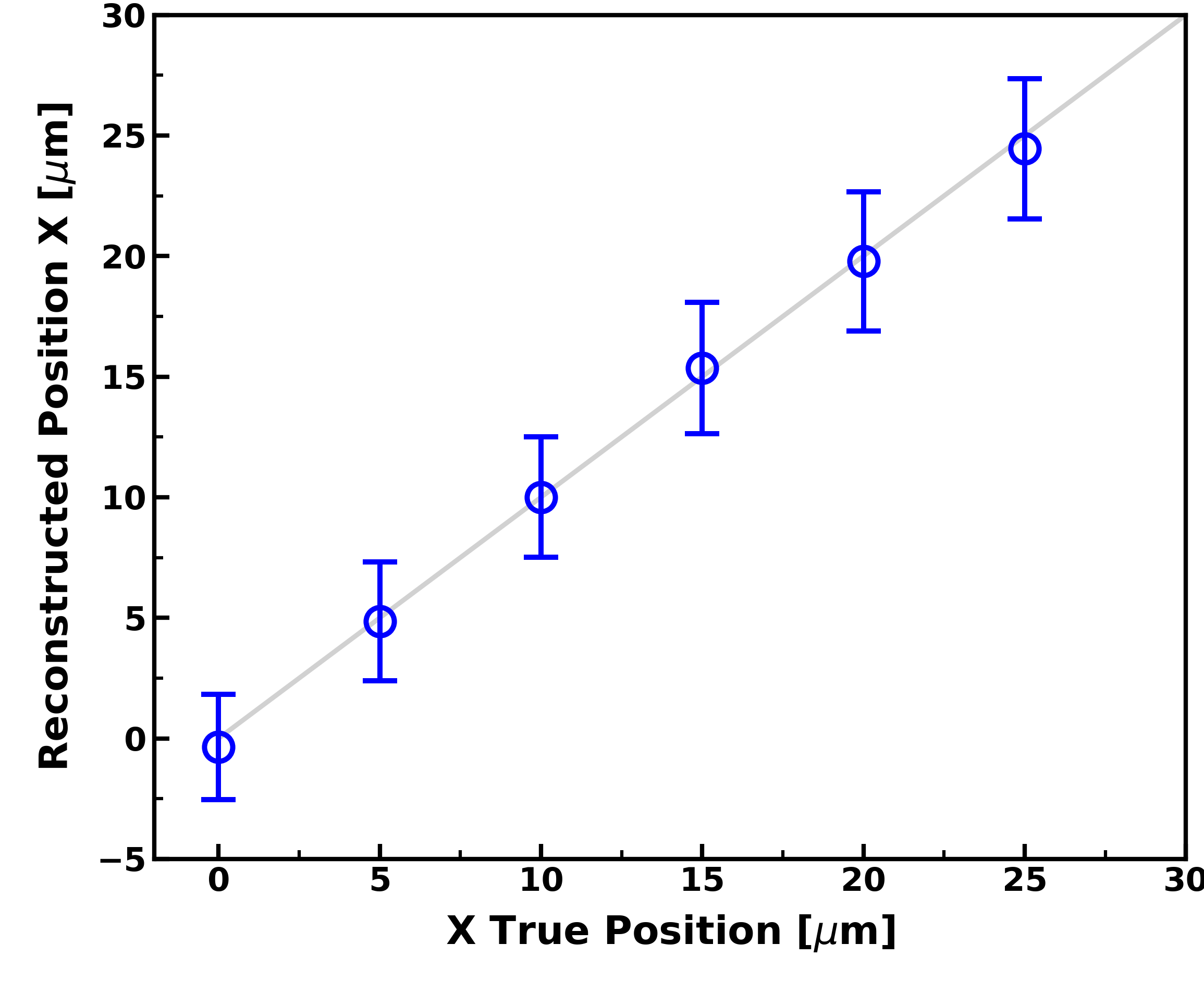}
}
\subfigure[]{
    \includegraphics[width=0.48\textwidth]{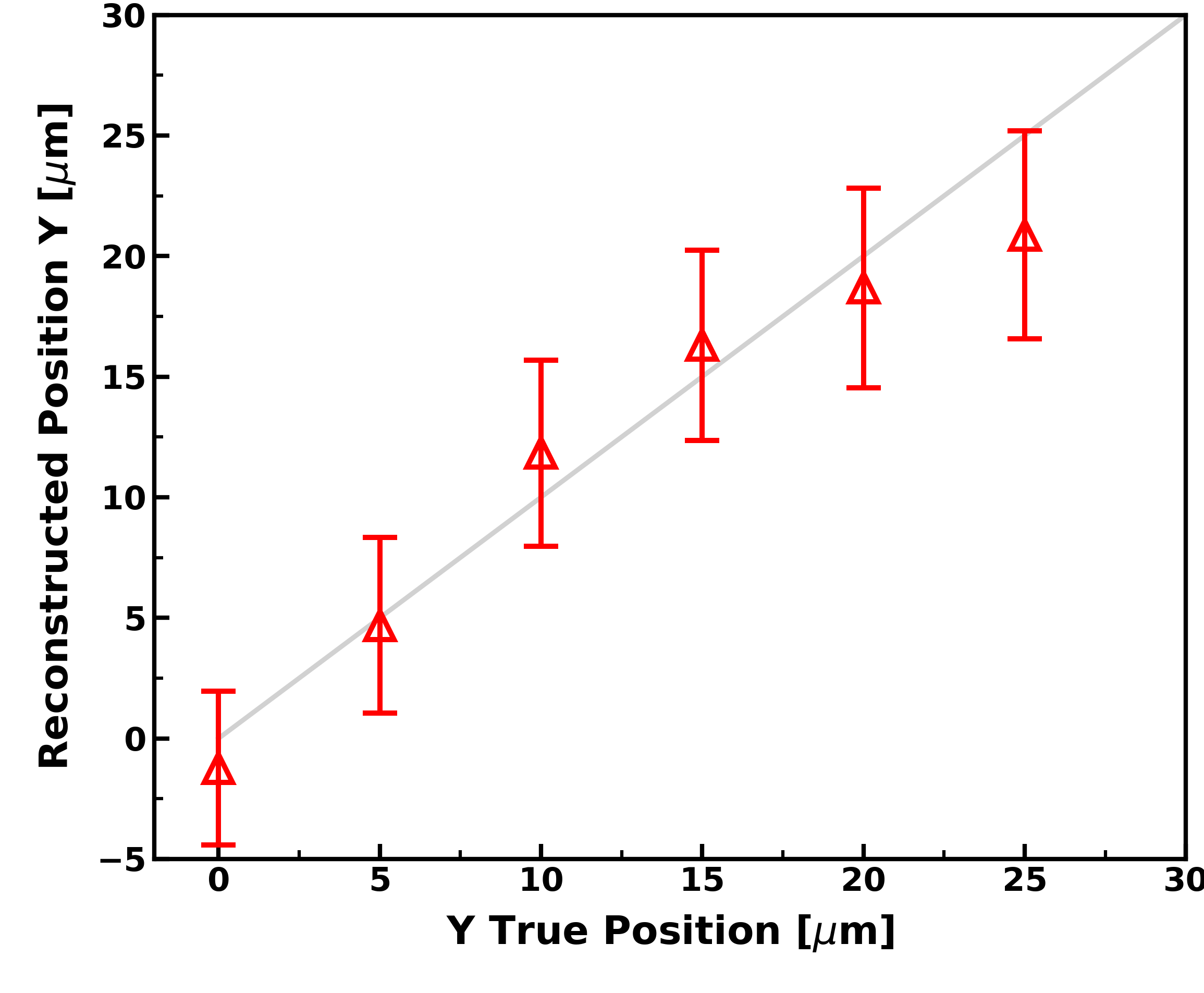}
}

\caption{\label{fig:timing}(a) Position reconstruction results in the $x$-direction utilizing strip1 and strip2 electrodes.
(b) Position reconstruction results in the $y$-direction utilizing strip3 and strip4 electrodes.}
\end{figure}

\begin{equation}
    R_x = \frac{Amp_2}{Amp_1 + Amp_2}, \quad R_y = \frac{Amp_4}{Amp_3 + Amp_4}
\end{equation}
Here, $Amp_i$ ($i=1, 2, 3, 4$) denotes the signal amplitude of four AC strip electrodes.

Fig. 7 shows the signal ratio $R$ exhibits a quasi-linear relationship with the particle hits position. Consequently, the reconstructed hit coordinates, $x$ and $y$, can be calculated using the following formula:
\begin{equation}
    x = \frac{R_x - c_x}{k_x}, \quad y = \frac{R_y - c_y}{k_y}
\end{equation}
\noindent Here, $R_x$ and $R_y$ represent the signal ratios, while $k$ and $c$ correspond to the slope and the intercept derived from the linear fitting of the ratio $R$.

Spatial resolution is defined as the standard deviation ($\sigma$) the difference between the true position and the reconstructed position. 
$1,000$ noise-added waveforms per hit were generated after two-stage amplification to serve as test samples. Fig. 8 shows the resulting $x$ and $y$ coordinate distributions reconstructed through the algorithm mentioned above.

Due to the overlap of AC strip 3 and 4 over AC strip 1 and 2, the signal coupling capability is deteriorated.
At a consistent noise level, this enhanced induction efficiency translates into a higher SNR for the bottom-layer strips. As a result, the reconstruction performance along the $x$-axis exhibits better linearity than along the $y$-axis.

Fig. 9(a) shows the histogram of $1,000$ reconstructed positions at $x = 5\,\mu\text{m}$ with an RMS noise level of $3.5\,\text{mV}$. 
And fitted by gauss, where $\sigma$ represents the local spatial resolution of the CS-LGAD with an $80\,\mu\text{m}$ pitch. 
By performing this statistical analysis at each incident coordinate, the position-dependent spatial resolution in both $x$ and $y$ directions for $3.5\,\text{mV}$ RMS noise is summarized in Fig. 9(b).

\begin{figure}[t]
    \centering
    \includegraphics[width=0.48\textwidth]{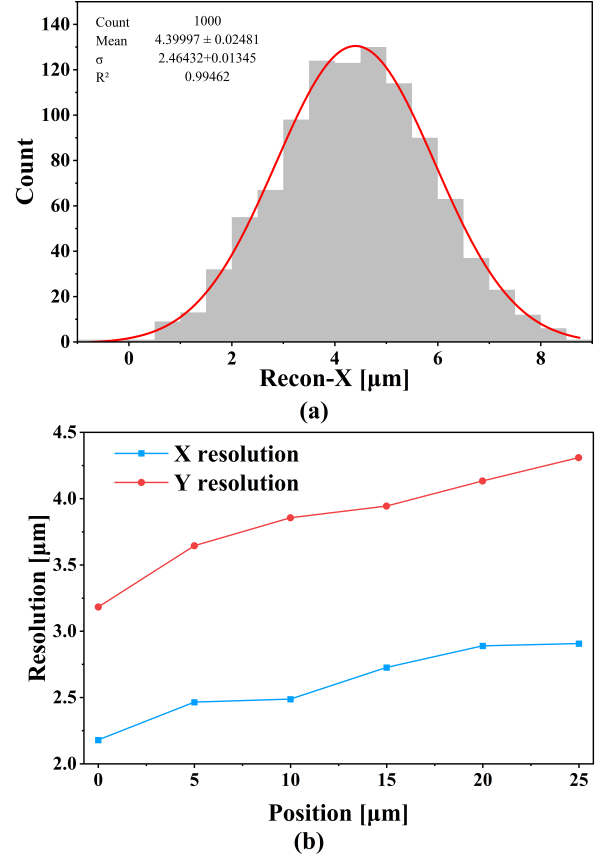}
    \caption{\label{fig:sigma}(a) Histogram of position reconstruction at $x = 5\,\mu\text{m}$; (b) Spatial resolution at an RMS noise level of $3.5\,\text{mV}$ (SNR $\approx 14$).}
\end{figure}

Fig. 9(b) shows at an RMS noise level of $3.5\,\text{mV}$ (SNR $\approx 14$), the CS-LGAD achieves a spatial resolution of better than $2.9\,\mu\text{m}$ in the $x$-direction and $4.4\,\mu\text{m}$ in the $y$-direction, they reached about 3.6\% and 5.5\% of the $80\,\mu\text{m}$ pitch size, respectively. Considering more unfavorable conditions, such as process non-uniformity and longer electrodes, the noise may increase and affect the spatial resolution. Therefore, the noise was increased from $3.5\,\text{mV}$ (SNR $\approx 14$) to $6\,\text{mV}$ (SNR $\approx 8$) to evaluate the spatial resolution under worse conditions. The results show that the resolutions in the $x$ and $y$ directions are $5\,\mu\text{m}$ and $7.5\,\mu\text{m}$, respectively. This indicates that the CS-LGAD can still provide good spatial resolution at a higher noise level.

This indicates that CS-LGAD also exhibits good spatial resolution when achieving simultaneous measurements in both x and y directions.

\section{Conclusion}
This work proposes a novel detector named AC-coupled CS-LGAD, which features overlapping strip AC-coupled electrodes and can be fabricated using a single-side process.
This design breaks through the limitations of existing strip-type AC-LGAD, achieving 2D position resolution on a single-layer sensor, while maintaining a lower readout channel density compared to pixel-type AC-LGAD.

{\frenchspacing At an RMS noise level of $3.5\,\text{mV}$ (SNR $\approx 14$), the simulated spatial resolution is better than $2.9\,\mu\text{m}$ in the x-direction and $4.4\,\mu\text{m}$ in the y-direction, corresponding to 3.6\% and 5.5\% of the $80\,\mu\text{m}$ pitch, respectively. As the RMS noise increases, the reconstructed position becomes more dispersed and the spatial resolution degrades. The degradation is more pronounced in the y-direction, mainly because the upper-strip coupling is weaker than the lower-strip coupling in the present geometry.}

Therefore, the proposed CS-LGAD holds great potential for particle position measurement, which is of significant importance for the development of time-trackers in future high-energy physics experiments.
In our next work, we will attempt to tape out a batch of CS-LGADs and evaluate their performance.

\section*{Acknowledgments}
This work was supported in part by the National Natural Science Foundation of China under Grant 12405224 and in part by the Guangdong Basic and Applied Basic Research Foundation under Grant 2024A1515140077.

\bibliographystyle{elsarticle-num}
\bibliography{CS_LGAD_REF}

\end{document}